\documentclass[11pt, a4paper, oneside, reqno]{amsart}
\usepackage{multirow}
\DeclareMathOperator*{\argmin}{arg\,min}
\makeatletter
\def\@settitle{\begin{center}%
		\baselineskip14\p@\relax
		\normalfont\LARGE\scshape\bfseries
		\@title
	\end{center}%
}
\makeatother

\makeatletter

\def\subsection{\@startsection{subsection}{2}%
	\z@{.5\linespacing\@plus.7\linespacing}{.5\linespacing}%
	{\normalfont\large\bfseries}}

\def\subsubsection{\@startsection{subsubsection}{3}%
	\z@{.5\linespacing\@plus.7\linespacing}{.5\linespacing}%
	{\normalfont\itshape}}

\usepackage[colorlinks	= true,
			raiselinks	= true,
			linkcolor	= MidnightBlue,
			citecolor	= Mahogany,
			urlcolor	= ForestGreen,
			pdfauthor	= {Chris van der Ploeg},
			pdftitle	= {},
			pdfkeywords	= {},   
			pdfsubject	= {},
			plainpages	= false]{hyperref}
\usepackage[usenames, dvipsnames]{color}
\usepackage{dsfont,amssymb,amsmath,subfig, graphicx,fancyhdr,mdframed}
\usepackage{amsfonts,dsfont,mathtools, mathrsfs,amsthm,wrapfig} 
\usepackage[]{siunitx}
\usepackage{ragged2e}
\usepackage{enumitem}
\usepackage{appendix}

\allowdisplaybreaks
\usepackage{algorithm}
\usepackage[noend]{algpseudocode}

\allowdisplaybreaks
\date{\today}

\patchcmd\@setauthors
  {\MakeUppercase{\authors}}
  {\authors}
  {}{}

\theoremstyle{remark}

\theoremstyle{remark}

\theoremstyle{definition}

\newcommand{\R}{\mathbb{R}}

\newcommand{\EZ}[1][]{e}
\newcommand{\fEZ}[1][]{e}

\newcommand{\argmax}{\mathop{\mathrm{arg\,max}}}

\graphicspath{{images/}}

\title[]{Fault Isolation for the Ink Deposition Process in \\ High-End Industrial Printers}

\author[C.D. van Peijpe]{Casper van Peijpe}
\author[F. Ghanipoor]{Farhad Ghanipoor}
\author{Youri de Loore}
\author[P.J.B. Hacking]{Pim Hacking}
\author[N. van de Wouw]{Nathan van de Wouw}
\author[P. {Mohajerin Esfahani}]{Peyman {Mohajerin Esfahani}}

\thanks{
The first and second authors contributed equally to this work. The authors are with the Delft Center for Systems and Control, Delft University of Technology ($\{${\tt C.D.vanPeijpe, P.MohajerinEsfahani$\}$@tudelft.nl}), the Department of Mechanical Engineering, Eindhoven University of Technology,
($\{${\tt F.Ghanipoor, N.v.d.Wouw$\}$@tue.nl}) and Canon Production Printing ($\{${\tt Youri.deLoore, Pim.Hacking$\}$@cpp.canon}).}

\begin{document}
\maketitle
\begin{abstract}   

This paper presents a mathematical framework for modeling the dynamic effects of three fault categories and six fault variants in the ink channels of high-end industrial printers. It also introduces a hybrid approach that combines model-based and data-based methods to detect and isolate these faults effectively. A key challenge in these systems is that the same piezo device is used for actuation (generating ink droplets) and for sensing and, as a consequence, sensing is only available when there is no actuation. The proposed Fault Detection (FD) filter, based on the healthy model, uses the piezo self-sensing signal to generate a residual, while taking the above challenge into account. The system is flagged as faulty if the residual energy exceeds a threshold. Fault Isolation (FI) is achieved through linear regression or a k-nearest neighbors approach to identify the most likely fault category and variant. The resulting hybrid Fault Detection and Isolation (FDI) method overcomes traditional limitations of model-based methods by isolating different types of faults affecting the same entries (i.e., equations) in the ink channel dynamics. Moreover, it is shown to outperform purely data-driven methods in fault isolation, especially when data is scarce. Experimental validation demonstrates superior FDI performance compared to state-of-the-art methods.

\end{abstract}

\newlength\figH
\newlength\figW
\setlength{\figH}{7cm}
\setlength{\figW}{12cm}	

\section{Introduction}
Over the past decades, fault diagnosis has been a focal point of research because of its vital role in maintaining the safety and reliability of engineering systems \cite{chen2012robust, ding2008model}. In this paper, the systems of interest are the printers of Canon Production Printing (CPP). CPP is a Canon subsidiary company located in Venlo, the Netherlands, that develops, manufactures and sells high-end industrial printers \cite{Canon}. Accurate fault diagnosis for the ink channels of these printers is essential for the improving print quality and maintenance services. The ink channels are subject to different faults, such as empty ink channels, partial or complete blockage of the nozzle or drying of the local ink \cite{annurev-fluid-022321-114001,Reinten_Jethani_Fraters_Jeurissen_Lohse_Versluis_Segers_2023}. All of these faults are examples of anomalies that would degrade image and printing quality. 
\begin{figure}[t]
    \centering
    \includegraphics[width=.9\textwidth]{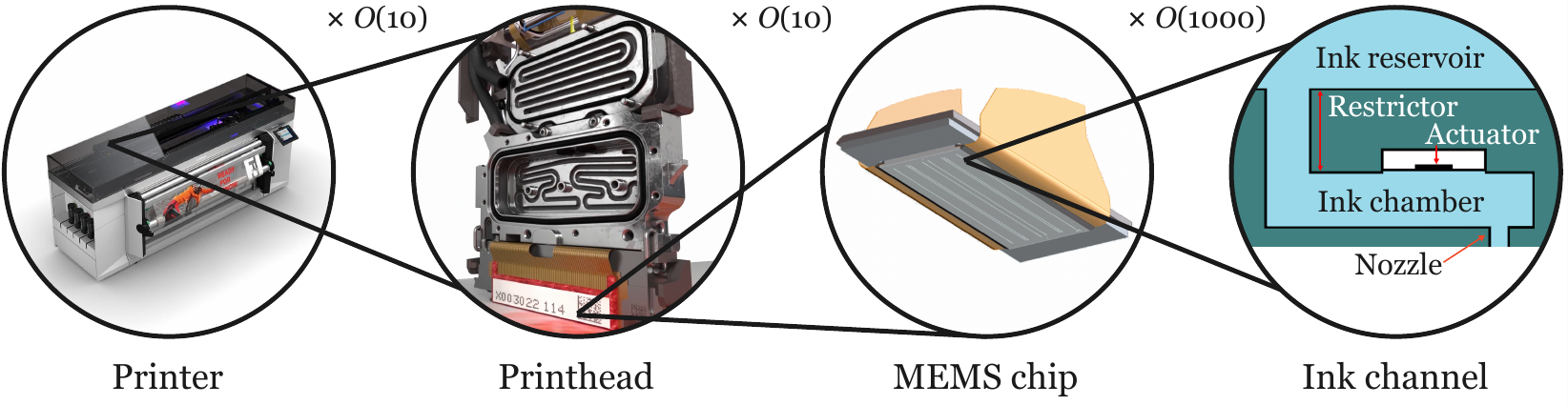}
    \caption{Decomposition of the printer where each component shown is smaller than and inside of the component to its left.}
    \label{fig:inkchannel}
\end{figure}
Preventing a low-quality print has several advantages. Firstly, improving the overall print quality results in satisfied customers and possibly attracts new businesses. Secondly, avoiding poor prints can reduce the need for reprints, saving ink, paper, and running costs associated with customer complaints and repairs. Additionally, reducing waste ink and paper is not only cost-effective, but it is also environmentally responsible. For these reasons, it is beneficial to detect a fault in advance and isolate faults in their early stages to compensate for a faulty ink channel before it reduces printing quality.

The printers considered in this paper consist of various components to ensure their proper functioning. In this context, the focus is on the ink channels, with up to hundreds or thousands located in the printheads, depending on the printer. In Figure~\ref{fig:inkchannel}, a decomposition is shown that indicates where each component is located and how many of each component are inside the component to its left. See \cite{Colorado} for more detailed information about the printers and \cite{UVgel} for details of the printheads.

The problem of detecting and isolating faults in the ink channel of printers is challenging. Both the time scale and the dimension of the system are microscopic, with a typical piezo self-sensing signal being tens of microseconds long \cite{Reinten_Jethani_Fraters_Jeurissen_Lohse_Versluis_Segers_2023}, an actuation chamber of mere millimetres in length and drop ejection nozzles of just tens of micrometres in diameter \cite{annurev-fluid-022321-114001}. The combination of the small time and length scales results in very small physical quantities, making it hard to observe the state of the system in real life. A so-called piezo self-sensing signal can be obtained by using the piezo actuator (used to induce ink deposition) as a sensor. In this way, it can be used to obtain information about the state of the ink channel. It is important to note that the piezo must be actuated first before the sensing signal can be obtained \cite{simons,byung,mihailovic}. Furthermore, depending on the printer, a signal from each of more than 100,000 ink channels is collected in less than 100 ms \cite{Canon}. Dealing with those large amounts of data emphasises the need for a computationally efficient method for fault diagnostics. A further challenge regarding data collection is that the piezo self-sensing signal cannot be captured during actuation, as the piezo can only serve as the actuator or the sensor at any point in time. Hence, input-output data is not available at the same time to support a solution for fault diagnostics. This also means that the state of the system cannot be assumed to be zero at the start of the sensing period since the system is being actuated before measurement acquisition starts. However, in the literature on fault diagnosis, the state of the system when measurement data is available is often assumed to be zero \cite{peyman,FRISK2002137,dong2024robust}, which is not the case here.

The frequency content of the self-sensing signals generated by the piezo actuator can be used for Fault Detection and Isolation (FDI) through signal-based fault tree methods with tunable thresholds, which is currently being used by the manufacturer for the health monitoring of their printers. This state-of-practice approach can be seen as a data-based method \cite{Ghasemzadeh2011, Hacking_2019}, which does not directly take into account any physics-based model of the system (ink channel dynamics). Besides, the thresholds used within this method are manual margins designed based on experience, making the method non-systematic in nature and hard to generalize to different printer types. 

Motivated by the aforementioned challenges, a hybrid model- and data-based FDI approach is proposed to detect and isolate faults using piezo self-sensing signals. This approach consists of two stages: Fault Detection (FD), which is the model-based part of the proposed methodology, and Fault Isolation (FI), which is the data-based part. The proposed method for FD builds upon previous results \cite{peyman, FRISK2002137}, where the task of FD involves generating a diagnostic signal sensitive to the occurrence of faults, known as the residual. This task is accomplished by designing a model-based filter fed by the piezo self-sensing signal. The filter formulation is modified to account for the system actuation and sensing data not being available simultaneously. For FI, inspired by \cite{van2022multiple}, an isolation filter is designed using a regression operator, as well as a classification algorithm, both using the output of the FD filter (i.e., residual signal). Contrary to the previous work, (simulated) faulty data is assumed to be available. Additionally, the FI filter design is adjusted to ensure that it can appropriately deal with scenarios in which only one fault can be present at a time (as is the case for the industrial printing use case in this paper). The main contributions of this paper are the following:
\begin{enumerate} [label=(\alph*)]
    \item \textbf{Modeling printer (ink channel) dynamics and possible faults:} We model the dynamics of ink channels for a class of industrial printers considering the fault categories of (1)~an empty ink channel, (2)~a blocked nozzle, and (3)~a dried nozzle. The modeling framework includes six different variants of these fault categories that are encountered in practice.

    \item \textbf{Methodologies for FDI:} In light of the practical limitation of actuation and sensor data being unavailable simultaneously, we develop a tailored FDI filter based on the model, in combination with available labeled experimental data. Moreover, the developed hybrid model-data fault diagnosis method addresses the limitation of conventional model-based approaches  by isolating different types of faults that affect the same entries in the ink channel dynamics \cite{cheng2015combined, dong2023robust, ghanipoor2025robust}. This capability is enabled by integrating data-driven classifiers, which effectively distinguish between different patterns in the residual generated by the model-based fault detector filter. Additionally, it is shown that the method outperforms purely data-driven approaches in fault isolation. 

    \item \textbf{Experimental validation on an industrial printer \cite{Canon}:} The results are validated experimentally, indicating that the proposed method provides superior performance for both fault detection and isolation compared to the current state-of-the-art method.

\end{enumerate}

The content of the remainder of paper is as follows. The problem statement of the printing use case is described in Section~\ref{section:SOPD}. A solution approach to the problem of fault detection and isolation is outlined in Section~\ref{section:solapp}. The results of this approach with simulated and experimental data are shown in Section~\ref{section:results}. Lastly, the conclusions that can be drawn from the results are presented, together with recommendations for future work in Section~\ref{section:conclusion}.
\vspace{-\baselineskip}  

\section{Printer Dynamics, Faults, and the Diagnosis Problem}
\label{section:SOPD}
\label{subsection:usecase}
The ink channel dynamics of the printer are presented in Section~\ref{subsection:dynamics}, the possible operational faults are explained in Section~\ref{subsectio:faults} and the problem of fault detection and isolation is described in Section~\ref{subsec:FDprob}.

\subsection{Mathematical model for ink channel dynamics}
\label{subsection:dynamics}

The smallest component in Figure \ref{fig:inkchannel}, the ink channel, is the subject of this study. When printing, a piezo-electric actuator is (dis)charged, inducing a mechanical deformation which displaces the ink from the ink chamber to both the restrictor and nozzle; see the sub-figure on the right side of Figure~\ref{fig:inkchannel}. A strong enough actuation will lead to a droplet being jetted from the nozzle onto the paper \cite{KWON200775}. The movement of the ink is captured in a four-dimensional state vector, denoted by $x$, which contains the volumes and flow rates displaced through the restrictor and the nozzle when actuating. The single variable voltage discharging the actuator is captured by the input $u$.

\begin{figure}[t]
    \centering
    \includegraphics[width=0.65\textwidth]{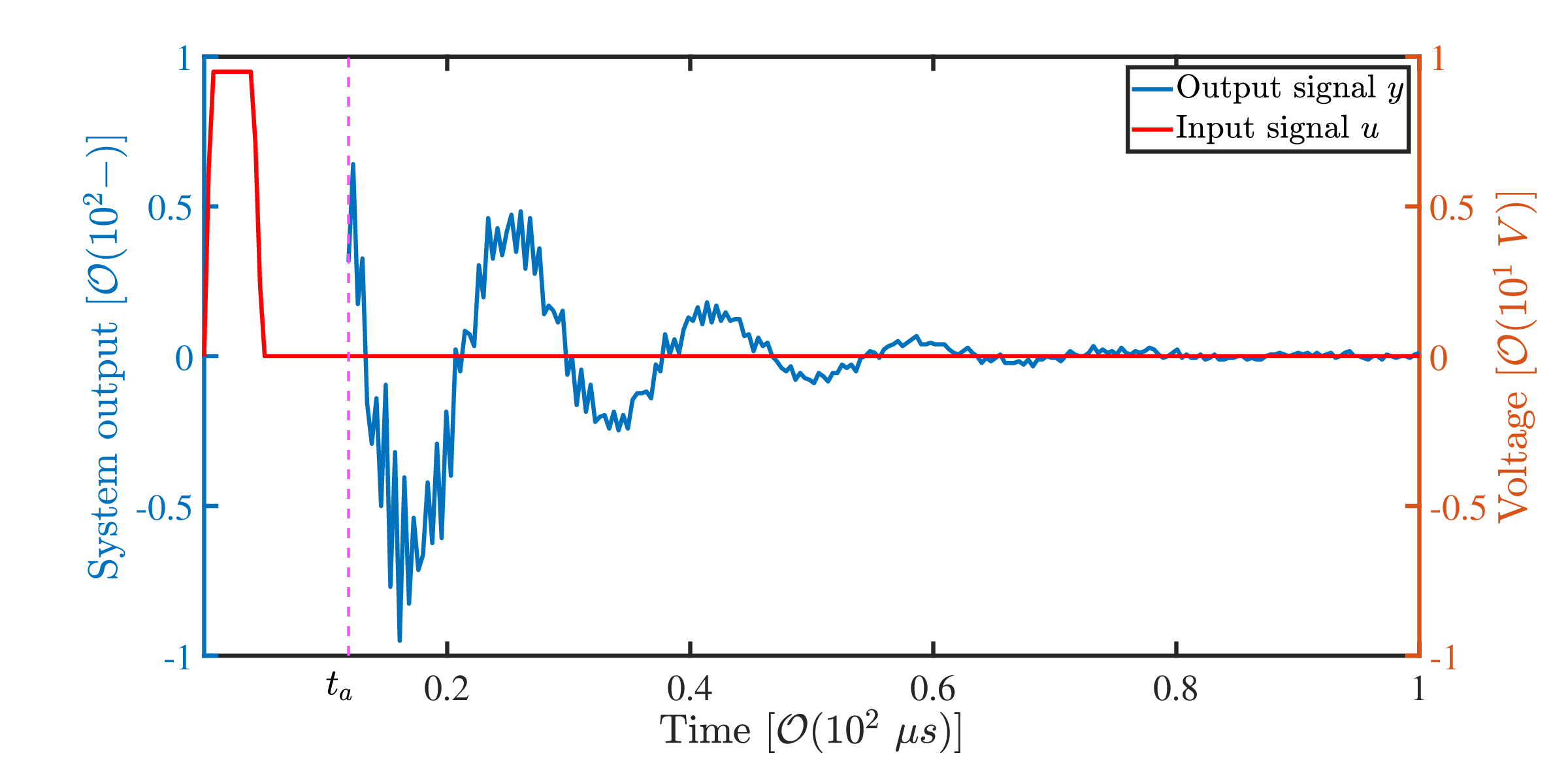}
    \caption{Typical input (actuation) signal and corresponding output (sensing) signal, with the output starting after the completion of the input at time $t_a$.}
    \label{fig:inputoutput}
\end{figure}

Following the fluid dynamics principles described in \cite{Reinten_Jethani_Fraters_Jeurissen_Lohse_Versluis_Segers_2023}, complemented by the addition of resistance in \cite{annurev-fluid-022321-114001}, the ink channel dynamics in the healthy case can be described as follows:
\begin{equation}
\label{eq:helmholtz}
\begin{aligned}
        \underbrace{\begin{bmatrix}
            \dot{V}_r \\
            \dot{V}_n \\
            \ddot{V}_r \\
            \ddot{V}_n 
        \end{bmatrix}}_{\dot{x}}&=
        \underbrace{\begin{bmatrix}
            0 & 0 & 1 & 0 \\
            0 & 0 & 0 & 1 \\
            -\frac{1}{I_r \beta_t} & -\frac{1}{I_r \beta_t} & -\frac{R_r}{I_r} & 0 \\
            -\frac{1}{I_n \beta_t} & -\frac{1}{I_n \beta_t} & 0 & -\frac{R_n}{I_n} 
        \end{bmatrix}}_A
        \underbrace{\begin{bmatrix}
            V_r \\
            V_n \\
            \dot{V}_r \\
            \dot{V}_n 
        \end{bmatrix}}_x +
        \underbrace{\begin{bmatrix}
            0 \\
            0 \\
            \frac{b}{I_r \beta_t} \\
            \frac{b}{I_n \beta_t}  
        \end{bmatrix}}_{B_u} u, \quad y = \underbrace{\begin{bmatrix}
            0&0&c&c  
        \end{bmatrix}}_C  \underbrace{\begin{bmatrix}
            V_r \\
            V_n \\
            \dot{V}_r \\
            \dot{V}_n 
        \end{bmatrix}}_x,
            \end{aligned}
\end{equation}
where $V_r$ and $V_n$ are the volumes of ink displaced through the restrictor and the nozzle when actuating, respectively. The parameters $I_r$ and $I_n$ are the inertances in the restrictor and the nozzle, respectively, $R_r$ and $R_n$ are the resistances in the restrictor and the nozzle, respectively, and $\beta_t$ is the total compliance of the system. The input $u$ is the voltage applied to the actuator. The actuator constant $b$ captures the relation between the voltage input and the influence on the channel pressure, which contributes proportionally to the time-derivative of the flow rate in the restrictor and nozzle via the factors $\frac{b}{I_r\beta_t}$ and $\frac{b}{I_n\beta_t}$, respectively. Acquisition constant $c$ captures the relationship between the flow rates (of ink displaced through the restrictor and the nozzle) in the ink chamber and the system output $y$. Note that a few microseconds after each actuation, the piezoelectric actuator can be used as a sensor to measure $y$ in \eqref{eq:helmholtz}. For a more detailed explanation of the model parameters, see \cite{Reinten_Jethani_Fraters_Jeurissen_Lohse_Versluis_Segers_2023}. A typical example of the process trajectories of both actuation and sensor acquisition is plotted in Figure \ref{fig:inputoutput}. 
Since the actuator acts as a sensor after actuation, the system is treated as autonomous with the non-zero state $x_a(t_a)$ (called the the acquisition state) at starting time $t_a$ of the acquisition. As a result, the term related to $B_u$ can be neglected in \eqref{eq:helmholtz} after $t=t_a$. 


\subsection{Operational faults} 
\label{subsectio:faults}
The operational faults studied in this article are caused by changes in some parameters of our mathematical model in \eqref{eq:helmholtz}. These faults are naturally of a multiplicative nature with the internal states of the system $x$. With this in mind, we can model each fault effect on the printer dynamics as follows:
\begin{equation}
\label{eq:generalsys}
\tag{{\rm ODE}($i$)}
\begin{cases} \dot{x}(t) & =(A+ B_{f_i} f_i) x(t), \quad i\in\{1,2,\ldots, n_f\}\\ 
y(t) & =C x(t),\end{cases}
\end{equation}
where the matrices $A$ and $C$ are as defined in \eqref{eq:helmholtz} and the signal $f_i \in \R_+$ is the $i$-th fault signal taking values ``0" whenever the system is healthy (i.e., no faults) or nonzero values when the fault occurs. Herein, $n_f$ denotes the number of fault variants, which are listed below and equal six. It is practically reasonable to assume that at most one $i$-th fault occurs at a time. Consequently, \eqref{eq:generalsys} captures the dynamics representing this sole $i$-th fault scenario. For every fault, the term $B_{f_i} f_i$ is added to $A$, where $B_{f_i}$ captures the changes in the parameters of $A$. The matrix $B_{f_i}$ may be different for every fault scenario and contains parameter changes related to that fault. Therefore, matrix $B_{f_i}$ is different for every fault $i$ and generally can be described as 
\begin{equation*}
    B_{f_i} = \begin{bmatrix}
            0 & 0 & 0 & 0 \\
            0 & 0 & 0 & 0 \\
            \delta (\frac{1}{{I_r \beta_t}})_i & \delta (\frac{1}{{I_r \beta_t}})_i & \delta (\frac{R_r}{I_r})_i & 0 \\
            \delta (\frac{1}{{I_n \beta_t}})_i& \delta (\frac{1}{{I_n \beta_t}})_i & 0 &\delta (\frac{R_n}{I_n})_i 
        \end{bmatrix}.
\end{equation*}
Note that in practice, since we do not know the actual entries in the matrix $B_{f_i}$, we model it as matrices with zero and nonzero elements to indicate which equations are affected. These are provided below, along with the fault scenarios:
\begin{description}
    \item[1) Empty ink channel (EC):] Refers to a fault in which an ink channel is not filled with ink. Every ink channel starts empty and subsequently it is filled with ink. An ink channel can also be fully emptied when printing. An empty channel can be modelled as a decrease in inertance of both the restrictor and nozzle, $I_r$ and $I_n$, causing the sensing signal $y$ to oscillate at a higher frequency. The corresponding fault matrix is denoted as $B_{f_1}$ and all non-zero elements in the general fault matrix $B_{f_i}$ are non-zero in $B_{f_1}$ as well. 
    \item[2) Blocked nozzle (BN):] Refers to a fault in which a dirt particle is present at the outlet of the nozzle. Such a particle blocks a certain percentage of the nozzle's cross-sectional area. This causes the nozzle inertance, $I_n$, and resistance, $R_n$, to increase, resulting in a faster decay of the sensing signal $y$. Two different types of blocked nozzle can be distinguished:
        (i) Fully blocked nozzle (FBN): Refers to a fault in which a nozzle is fully blocked, implying no printing can be done;
        (ii) Partially blocked nozzle (PBN): Refers to a fault in which a part of the nozzle is blocked, implying ink exits the nozzle incorrectly.
    The corresponding fault matrices are denoted as $B_{f_{2}}$ and $B_{f_{3}}$, for the FBN and PBN, respectively. Both matrices have the same non-zero elements, which are $ B_{f_2}(4,1) $, $ B_{f_2}(4,2) $, $ B_{f_2}(4,4) $, and similarly $ B_{f_3}(4,1) $, $ B_{f_3}(4,2) $, $B_{f_3}(4,4)$. 
    \item[3) Dried nozzle (DN):] Refers to a fault in which the ink at the nozzle's outlet has evaporated, losing some of its most volatile components, such as water. This causes the viscosity to increase and therefore the resistance $R_n$ to locally increase, leading to a faster decay of the sensing signal $y$. Three different drying levels are distinguished:
        (i)~Slightly dried nozzle (SDN),
        (ii)~Intermediately dried nozzle (IDN),
        (iii)~Deeply dried nozzle (DDN).
    The corresponding fault matrices are denoted as $B_{f_{4}}$, $B_{f_{5}}$, and $B_{f_{6}}$, for the SDN, IDN, and DDN, respectively. All three matrices have the same non-zero element, which is $ B_{f_4}(4,4) $ and similarly $ B_{f_5}(4,4)$ and $ B_{f_6}(4,4) $. 
   
\end{description}

The problem to be solved in this paper can now be stated.
\subsection{Fault detection and isolation problem}
\label{subsec:FDprob}
We aim to design an FD filter that can distinguish between healthy and faulty systems using the output data $y(t)$ and an approximated acquisition state $\hat{x}_a$. The FD filter generates a residual signal~$r$ that indicates whether the measurements are healthy or are affected by the fault signal $f_i$. Formally, the residual can be represented as a function $r(x_a, f_i)$, and the FD design is ideally defined by the following mapping requirements: 
\begin{subequations} \label{eq:map}
\begin{align}
\label{eq:mapA} 
& x_a \mapsto r(x_a, 0) \equiv 0, \\
\label{eq:mapB}
& f_i \mapsto r(x_a, f_i) \neq 0, \quad \forall i\in\{1,2,\ldots, n_f\}.
\end{align}
\end{subequations}

For fault isolation, conventional model-based methods, such as \cite{peyman}, use filter banks, with each filter sensitive to one fault and insensitive to others. However, this approach fails when faults share the same system dynamics entries. Isolation relies on amplifying one fault's signal while suppressing another's in the same residual \cite{cheng2015combined, dong2023robust}, which is impossible when both faults affect the same entry. To address this challenge, we aim to design a fault isolation (FI) filter capable of isolating the occurring fault using the output of the fault detection (FD) filter, $r$, building on the results in \cite{van2022multiple}. The FI filter is trained offline using available labeled experimental data for each fault scenario. In the online implementation, the filter generates a vector $\varphi^{\star}$ (see the output of right block in Figure \ref{fig:blockscheme}), where each row represents the percentage of occurrence for each fault scenario, and the fault with the highest percentage is considered to be the occurring fault.

\section{Fault detection and isolation approach}
\label{section:solapp}
\begin{figure}[t]
    \centering
    \includegraphics[width=0.85\textwidth]{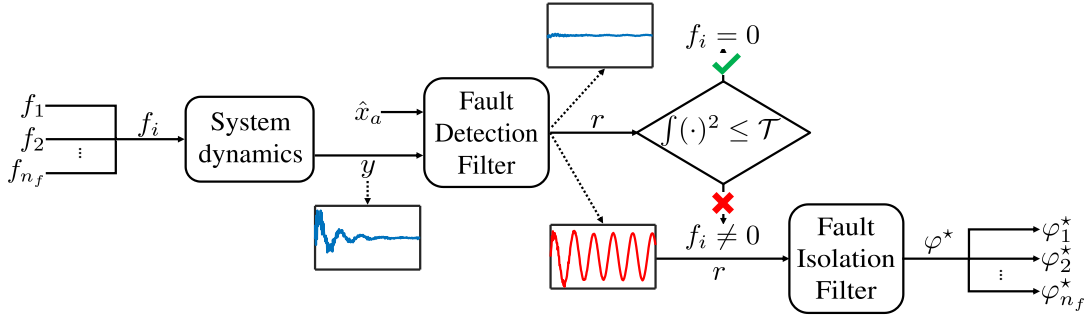}
    \caption{Overview of the proposed FDI scheme.}
    \label{fig:blockscheme}
\end{figure}

In this section, the proposed solution is presented. Figure \ref{fig:blockscheme} illustrates an overview of the proposed FDI scheme, where different types of faults $f_i$ entering the system are detected and isolated. The methods for fault detection and fault isolation are provided in Sections~\ref{subsection:FD} and \ref{subsection:FI}, respectively.

\subsection{Fault detection filter}
\label{subsection:FD}
As a preliminary step, an accurate model is obtained using an off-the-shelf grey-box system identification method in MATLAB, namely extended least squares for parameter estimation \cite{greyest}. This method uses healthy sensor data to estimate system parameters in \eqref{eq:helmholtz} via solving the following optimization problem:
\begin{equation}
\label{eq:sysID}
    \{ A,C,\hat{x}_a \} = \argmin_{A,C,\hat{x}_a} \int_{t_a}^{t_f}  | y(t)-Ce^{At}\hat{x}_a | dt,
\end{equation}
where $t_f$ denotes the final time of the signal $y$ collected from the printer. The optimization problem is solved using the interior-point algorithm \cite[Ch.~11]{boyd}.
The resulting matrices $A$ and $C$ and the estimated acquisition state $\hat{x}_a$ are then used to develop a model-based fault detection filter. 

To construct the FD filter, first, the state-space representation of the system in \eqref{eq:generalsys} is converted from time domain to the Laplace domain as follows:
\begin{equation}\label{eq:ode_laplace}
\begin{cases} sX(s)-x_a & =A X(s) + B_{f_i} \Psi_i(s),\\ Y(s) & =C X(s),\end{cases}
\end{equation}
where, $s \in \mathbb{C}$ is the complex Laplace variable used in the Laplace transform. $X(s), Y(s)$ and $\Psi_i(s)$ are the Laplace transforms of $x(t), y(t)$ and $f_i x(t)$, respectively. Equation \eqref{eq:ode_laplace} can be written as
\begin{equation}
\label{eqDAE}
\tag{{\rm DAE}($i$)}
\begin{aligned}
&\underbrace{\left[\begin{array}{cc}
-sI+A  \\
C 
\end{array}\right]}_{H(s)=H_0+H_1s}X(s)+ \left[\begin{array}{cc}
x_a  \\
0 
\end{array}\right]+
\underbrace{\left[\begin{array}{cc}
0  \\
-I 
\end{array}\right]}_{L(s)}Y(s) 
+
\underbrace{\left[\begin{array}{cc}
B_{f_i}  \\
0 
\end{array}\right]}_{F_i(s)}\Psi_i(s)
=0,
\end{aligned}
\end{equation}
where $H(s) \in \mathbb{C}^{(n+m)\times n}$, $L(s) \in \mathbb{C}^{(n+m)\times m}$, and $F_i(s) \in \mathbb{C}^{(n+m)\times n}$, for all $i \in\{1,2,\ldots, n_f\}$. This system representation is also known as the Differential-Algebraic Equation (DAE) (represented in the Laplace domain).


The FD filter can be stated as follows, generating the residual signal defined in the Laplace domain:
\begin{equation} \label{eq:residual}
    R(s) := \frac{N(s)L(s)}{\alpha(s)}Y(s) + \frac{N(s)}{\alpha(s)}\left[\begin{array}{cc}
\hat{x}_a  \\
0 
\end{array}\right],
\end{equation}
where $R \in \mathbb{C}$ is the residual in the Laplace domain, $\hat{x}_a$ denotes the approximated acquisition state from \eqref{eq:sysID}, $N(s)$ is designed to solve the problem stated in Section \ref{subsec:FDprob} (or more specifically to satisfy \eqref{eq:map}), and $\alpha(s)$ is chosen such that the transfer function of the residual generator $R(s)$ is (strictly) proper. The method used to design $\alpha(s)$ is described later.  This residual can be transformed to the time domain to form the signal $r$. To design $N(s)$, the residual in \eqref{eq:residual} must be expressed in terms of both state-dependent and fault-dependent terms. It follows from \eqref{eq:residual} and \eqref{eqDAE} that 
\begin{equation} \label{eq:residual_proof_end}
\begin{aligned}
    R(s) = & - \frac{N(s)H(s)}{\alpha(s)}X(s) - \frac{N(s)F_i(s)}{\alpha(s)}\Psi_i(s) 
+\frac{N(s)}{\alpha(s)}\left[\begin{array}{cc}
\hat{x}_a - x_a  \\
0 
\end{array}\right].
\end{aligned}
\end{equation}
To satisfy \eqref{eq:mapA}, the first term in \eqref{eq:residual_proof_end} must be zero (i.e., $N(s)H(s) = 0$). Similarly, to satisfy \eqref{eq:mapB}, the second term must be nonzero (i.e., $N(s)F_i(s) \neq 0$). It is shown in \cite{peyman} that these conditions on the transfer functions can be expressed in terms of matrices as follows:
\begin{subequations} \label{eq:map2}
\begin{align}
    \label{eq:mapA2} & \bar{N}\bar{H}=0, \\
    \label{eq:mapB2} & \bar{N}\bar{F}_i \neq 0, \quad \forall i\in\{1,2,\ldots, n_f\},
\end{align}
\end{subequations}
with $\bar{H} \in \mathbb{R}^{d_N(n+m)\times n(d_N+1)}$ as
\begin{equation*}
\Bar{H} :=
    \left. \left[ \begin{array}{cccc}
   H_0 &H_1& & \\
   &\ddots&\ddots&\\
   && H_0 & H_1
    \end{array} \right] \right\} d_N,
\end{equation*}
where, $H_0$ and $H_1$ as defined in \eqref{eqDAE}, $d_N$ is the order of the filter, $\bar{N} \in \mathbb{R}^{1\times {d_N(n+m)}}$, and $\bar{F_i} \in \mathbb{R}^{d_N(n+m)\times n d_N}$ as given by
\begin{equation*}
\Bar{F_i} :=
    \left. \left[ \begin{array}{cccc}
   F_i & & \\
   &\ddots&\\
   && F_i
    \end{array} \right] \right\} d_N,
\end{equation*}
where $F_i$ as defined in \eqref{eqDAE}. Generally, to create the FD filter, the  linear programming problem in \eqref{eq:map2} is solved \cite{peyman}. However, for this specific system (i.e., the ink channel system), the linear programming constraint \eqref{eq:mapA2} has only one solution, meaning solving the linear programming problem is not needed. An approach to design $\bar{N}$ is to calculate the left null space of $\bar H$ as 
$
    \Bar{N} = (\mathcal{N}(\Bar{H}^\top))^\top = \left[\begin{array}{llllll}
N_0 & N_1 & \cdots & N_{d_N}
\end{array}\right].
$

As for implementation purposes, the fault detection filter formulation should be derived in the Laplace domain, $\Bar{N}$ can be converted to $N(s)$ using 
$$
    N(s):=\sum_{i=0}^{d_N} N_i s^i.
$$

Note that clearly from \eqref{eq:residual_proof_end}, it can been verified that if $\hat{x}_a - x_a$ is small, the residual signal will only generate the second term, which is zero in the healthy case because $\Psi_i(s)$ (i.e., the Laplace transform of $f_i x(t)$) is zero, and non-zero when a fault occurs, as the condition in \eqref{eq:mapB2} imposes.

Due to errors caused approximation of the acquisition state (i.e., $\hat{x}_a - x_a$), modeling uncertainties, and measurement noise, a non-zero residual does not necessarily indicate that the system is faulty, which can lead to some healthy cases being misidentified as faulty. Therefore, to determine whether a residual signal is faulty, a threshold $\mathcal{T}$ on the energy of the residual signal, defined as the integral of the square of the signal, is established as 
\begin{equation} \label{eq:threshold}
    \mathcal{T} = \mu \max_{r_h \in {\mathcal{S}_h}} \left( \int_{t_a}^{t_f} r_h^2(t) \, dt \right),
\end{equation} 
where $\mathcal{S}_h$ is the healthy residual dataset which is produced by passing measurements $y$, including measurement noise, through the FD filter, $r_h$ represents the healthy residuals taken from this dataset, and $\mu \in \R_+$ is a user-determined factor designed to make the detection method more robust against false alarms caused by approximation errors, modeling uncertainties, and measurement noise (see Section \ref{sec:FD_result} for design examples). If this threshold is exceeded by the residual energy (i.e., $\int_{t_a}^{t_f} r^2(t) dt$), the system is classified as faulty (see Figure \ref{fig:blockscheme}). Besides this threshold, other metrics, such as the maximum value of the residual, which may be computationally faster, can also be used to determine whether a signal is faulty.

The poles of the filter, represented by the denominator $\alpha(s)$  in \eqref{eq:residual}, need to be selected carefully. In \cite{peyman}, the denominator of the FD filter is chosen as an arbitrary transfer function of sufficiently high order, with roots having negative real parts to ensure the transfer function of the filter is both proper and stable. However, it has been observed that the denominator can influence the performance of the filter for the system under consideration.

In this paper, since the energy of the residual is used for fault detection, the FD filter denominator is designed to amplify the energy of the residual, making faults easier to detect. This is achieved by placing two marginally stable poles of the FD filter at $s=\pm \omega_r i$, where $\omega_r$ is the desired residual frequency, which can be chosen arbitrarily. The frequency can be selected so that a desired number of oscillations occurs within a known time frame $T$. Since this pole pair does not have a negative real part, the residual does not converge to zero, and the energy continues to increase. This amplifies the presence of a fault, as a faulty residual results in a larger amplitude compared to a healthy residual. The remaining poles are chosen to be arbitrarily stable, as they do not influence the behavior of the residual significantly. Thus, the denominator can be expressed as follows:
\begin{equation*}
\alpha (s) = (s^2+\omega_r^2) \beta (s), \quad \omega_r = \frac{n_o}{T},
\end{equation*}
where $\beta(s)$ is a transfer function with roots with negative real parts of order $d_N-2$, and $n_o$ is the desired number of oscillations. 

Once a fault has been detected, the next step is to isolate the detected fault. Towards this goal, an FI filter is designed in what follows. 

\subsection{Fault isolation filter}
\label{subsection:FI}
\begin{figure}[t]
    \centering
    \includegraphics[width=0.85\linewidth]{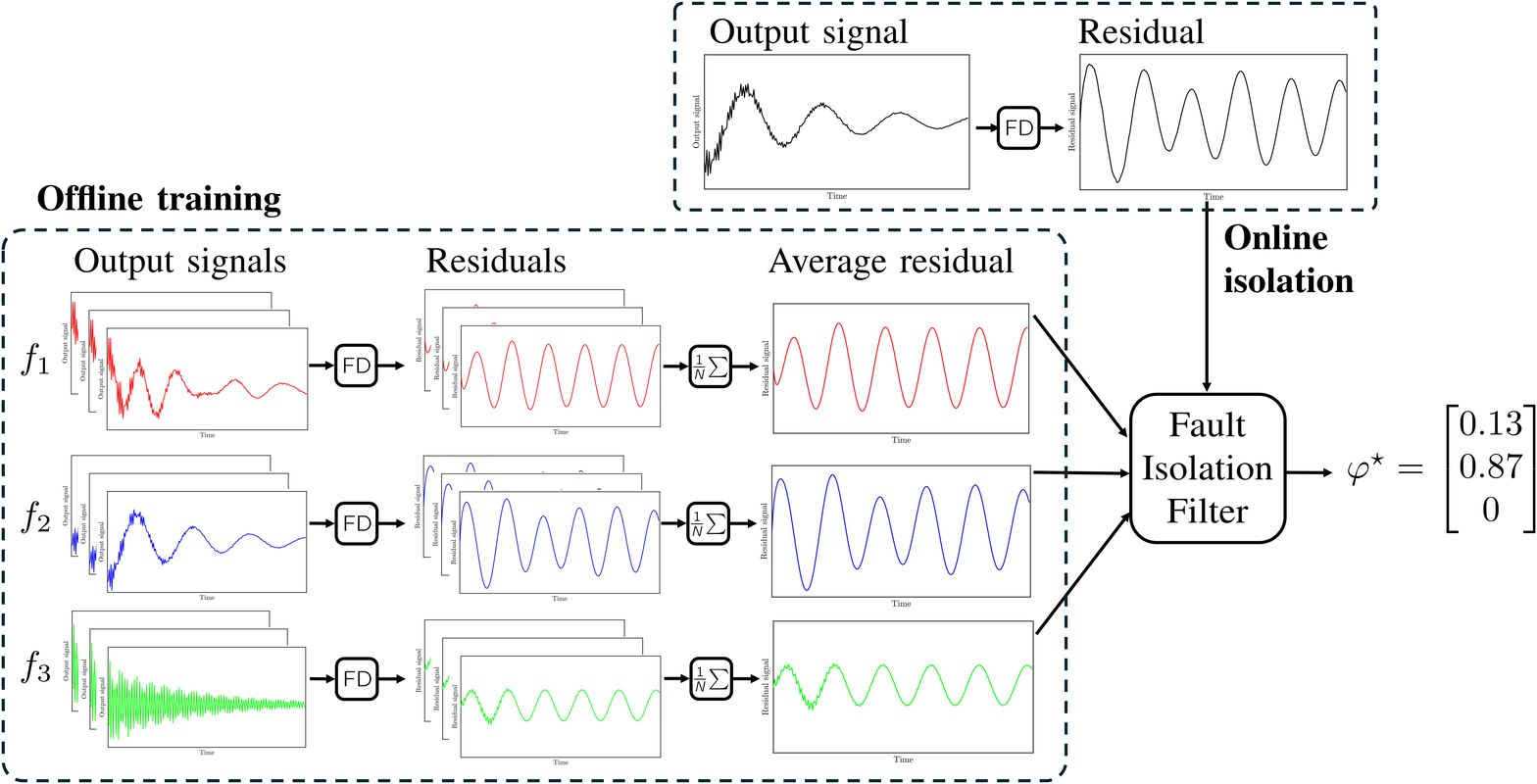}
    \caption{Schematic overview of the offline training (design) and online application of the FI filter.}
    \label{fig:FIscheme}
\end{figure}

The FI filter is inspired by \cite{van2022multiple}, where a Linear Regression (LR) model is also used for fault isolation. To train the FI filter for each fault, labeled faulty data must be available. These LR models are trained offline in a discrete setting. Therefore, for offline training, continuous residuals obtained from the proposed FD filter should be discretized, and the $N$ discretized residuals per fault class are averaged for each fault scenario as follows:
\begin{equation*}
 \rho_i[k] = \frac{1}{N} \sum_{j=1}^N r_i^j[k], \quad \forall i\in\{1,2,\ldots, n_f\}, k \in\{k_a,\ldots, k_f\},
\end{equation*}
where $\rho_i[k] \in \R$ represents the average residual for each fault class $i$ (see the offline training part in Figure \ref{fig:FIscheme}, where only three faults are shown for simplicity), and $k_a, k_f$ are the samples corresponding to $t_a, t_f$, respectively. Next, the matrix $R$, containing the averaged residuals for each fault class $\rho_i[k]$, can be constructed as follows:
\begin{equation*}
R = \begin{bmatrix}
{\rho}_1[k_a] & {\rho}_2[k_a] & \cdots & {\rho}_{n_f}[k_a] \\
{\rho}_1[k_a+1] & {\rho}_2[k_a+1] & \cdots & {\rho}_{n_f}[k_a+1] \\
\vdots & \vdots & \ddots & \vdots \\
{\rho}_1[k_f] & {\rho}_2[k_f] & \cdots & {\rho}_{n_f}[k_f]
\end{bmatrix}.\end{equation*} 
During the online isolation process, LR is used to identify which expected residual $\rho_i[k]$, obtained from offline training, is closest to the measured residual signal $r$ to be isolated (see the online isolation part in Figure \ref{fig:FIscheme}). The LR model quantifies the contribution of each fault as a vector $\varphi^\star \in \mathbb{R}_+^{n_f}$, which is the solution to the following optimization problem:
\begin{equation*}
\varphi^{\star}=\left\{\begin{array}{lll}
\argmin _{\varphi} & \left|\left| R\varphi-r\right|\right| & \\
\text { s.t. } & \textstyle \sum_{i=1}^{n_f}\varphi_i & = 1 ,\\
& \varphi_i & \geq 0, \quad \forall i\in\{1,2,\ldots, n_f\},
\end{array}\right.
\end{equation*} 
where 
\begin{equation*}
        (\varphi)^{\top} =  \left[\begin{array}{ccc} \varphi_{1} & \dots & \varphi_{n_f}  
    \end{array}\right],
\end{equation*}
is used as a regressor. The fault index $i^\star$ corresponding to the highest value in $\varphi^\star$ is considered as the isolated fault, i.e., $i^\star = \argmax_i {\varphi_i^\star}, i\in\{1,2,\ldots, n_f\}$ (for the simplified example illustrated in Figure \ref{fig:FIscheme}, $i^\star$ is 2). Instead of LR models, other classifiers, such as the k-nearest neighbors algorithm (KNN), can be used (see \cite{knn}). In this case, the residual signal is assigned to the class of its nearest neighbor. 
In the following sections, the FD and FI filters are evaluated using both simulation and experimental data concerning the ink channel dynamics in an industrial printer.


\section{Simulation and Experimental Results}
\label{section:results}
First, the structure of the data is analyzed in Section~\ref{subsection:data}. Then the FD and FI filters are tested for both the simulation and experimental data in Sections~\ref{subsection:resFD} and \ref{subsection:resFI}, respectively. 
\subsection{Data analysis}
\label{subsection:data}
\begin{figure}[b]
    \centering
    \includegraphics[width=0.63\textwidth]{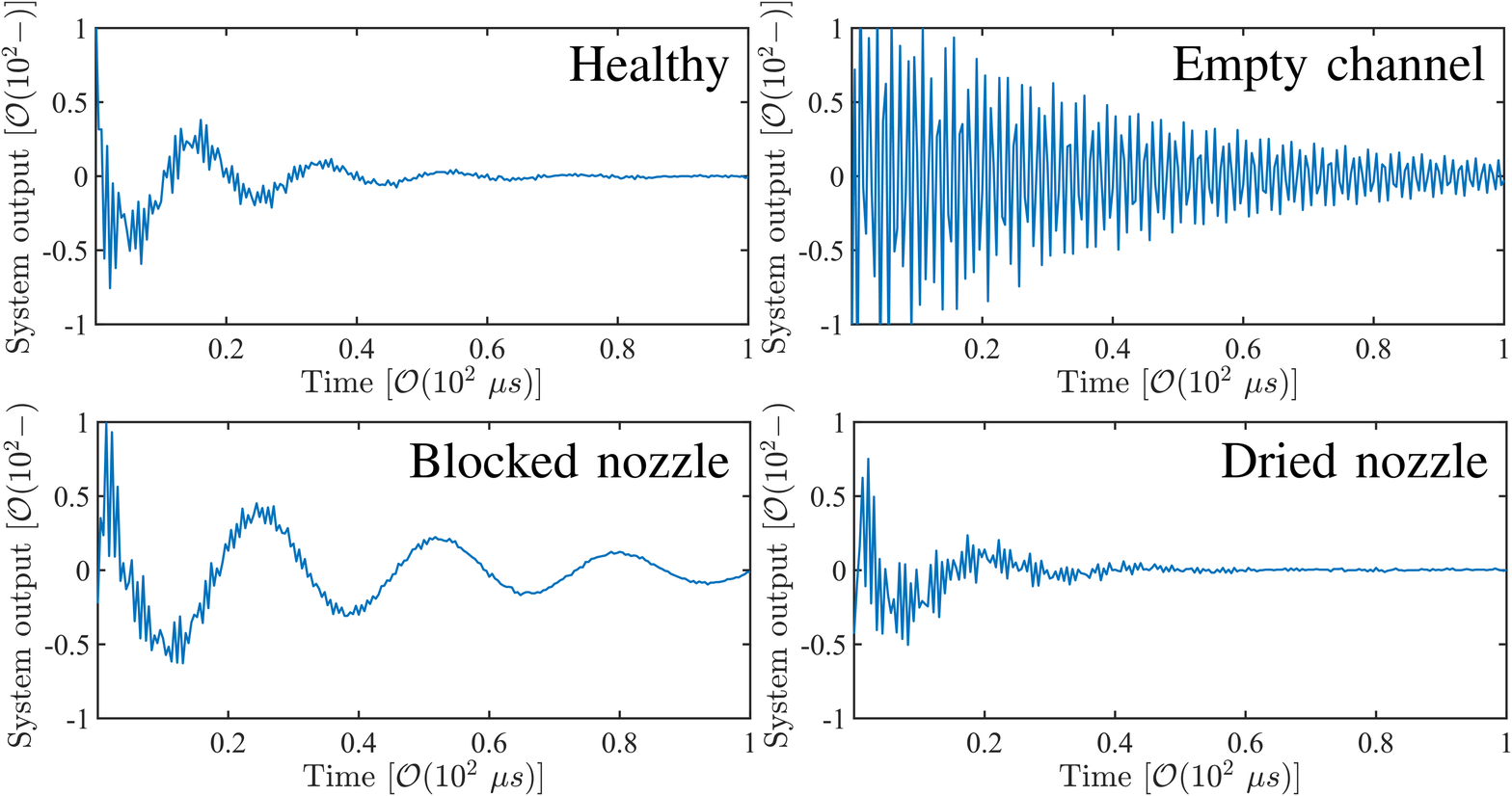}
    \caption{Examples of sensing signals $y$ for simulated fault categories (here $t_a=0$).}
    \label{fig:signal_examples}
\end{figure}

To evaluate both filters, simulated data and experimental data are used. Simulated data is generated based on knowledge of faults and the nature of the available experimental data. 3,825 output signals, $y$, are generated, where 2,475 are healthy signals and every fault class contains 225 signals. Hence, all of the six fault classes mentioned in Section \ref{subsection:usecase} are considered. One healthy example and three examples of faulty signals per fault category are illustrated in Figure \ref{fig:signal_examples}. 

\begin{figure}[b]
    \centering
    \includegraphics[width=0.63\textwidth]{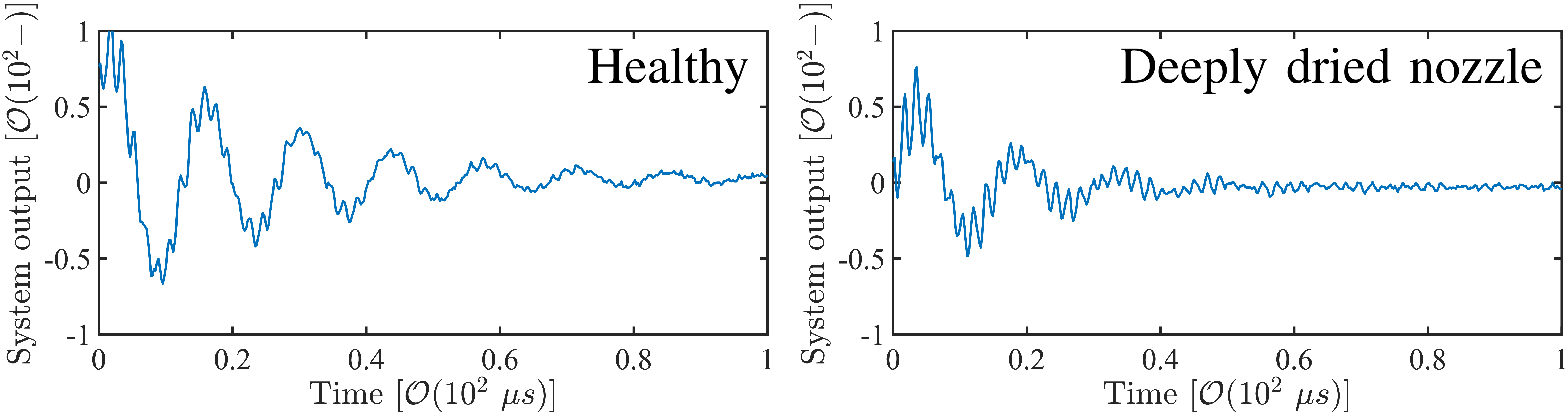}
    \caption{Examples of sensing signals $y$ for experimental fault categories (here $t_a=0$).}
    \label{fig:signal_examples_real}
\end{figure}

Experimental data is collected by purposely drying the nozzle. The procedure to overcome a drying nozzle, called prefiring (mixing the ink in the nozzle and ink chamber to bring the ink in the nozzles back to a usable viscosity), is performed to collect data for fault scenarios ranging from deeply dried to healthy. Three datasets are generated: the first, with 48 signals, represents healthy nozzles, where sufficient prefires remove all dried ink. The second, with 39 signals, represents intermediately dried nozzles, where some prefires partially remove dried ink. The third, with 60 signals, represents deeply dried nozzles, with no prefires performed. Two examples of measured faulty signals per fault category are illustrated in Figure \ref{fig:signal_examples_real}. Notably, separate datasets of the same length as the test data are used for the offline training of the FI filter for both the simulated and experimental data.

\subsection{Fault detection results} \label{sec:FD_result}
\label{subsection:resFD}
The preliminary system identification results, obtained using the off-the-shelf grey-box method in MATLAB (explained in Section \ref{subsection:FD}), are provided in the Appendix.

Some fault classes for the simulated dataset include signals that show characteristics similar to healthy signals. To prevent a large number of false positives, i.e., healthy signals being detected as faulty by the FD filter, the multiplication factor for the threshold in \eqref{eq:threshold} is chosen as $\mu = 1$.
For the experimental dataset, healthy and faulty signals can be distinguished more easily. Therefore, a multiplication factor of $\mu=2$ is chosen to prevent false positives for the experimental data. 

We present the FD results of the proposed method by comparing it with the State-of-Practice (SoP) approach. The SoP method relies on a signal-based fault tree approach with tunable thresholds, which are adjusted based on experience. For further details on the SoP method, refer to \cite{Ghasemzadeh2011, Hacking_2019}.

The FD results of the simulated data are summarized in Table~\ref{tab:CMFDsim}. The proposed FD filter achieves a Total Detection Rate (TDR) \cite{vanhybrid} of 97.6\%, surpassing the 93.9\% of the manufacturer's current method. Its False Alarm Rate (FAR) \cite{vanhybrid} is 0.04\%, compared to 0\% for the SoP approach. The proposed method has only one false negative, which is related to the slightly dried nozzle. Nevertheless, the substantial improvement in fault detectability halves the number of false detections, underscoring the superior reliability of the proposed approach compared to the SoP method. 

\begin{table}[t]
      \caption{FD Filter results for the simulated data.}
      \label{tab:CMFDsim}
      \centering
        \begin{tabular}{ll|ll|}
\cline{3-4}
 &  & \multicolumn{2}{c|}{FD filter} \\ \cline{3-4} 
 &  & \multicolumn{1}{l|}{Healthy} & Faulty \\ \hline
\multicolumn{1}{|c|}{\multirow{2}{*}{Ground truth}} & Healthy & \multicolumn{1}{l|}{2474} & 1 \\ \cline{2-4} 
\multicolumn{1}{|c|}{} & Faulty & \multicolumn{1}{l|}{92} & 1258 \\ \hline
\end{tabular}
\end{table}

\begin{table}[b]
      \centering
        \caption{FD Filter results for the experimental data.}    
        \label{tab:CMFDreal}
        \begin{tabular}{ll|ll|}
\cline{3-4}
 &  & \multicolumn{2}{c|}{FD filter} \\ \cline{3-4} 
 &  & \multicolumn{1}{l|}{Healthy} & Faulty \\ \hline
\multicolumn{1}{|c|}{\multirow{2}{*}{Ground truth}} & Healthy & \multicolumn{1}{l|}{48} & 0 \\ \cline{2-4} 
\multicolumn{1}{|c|}{} & Faulty & \multicolumn{1}{l|}{0} & 99 \\ \hline
\end{tabular}
\end{table}

For the experimental data, the proposed FD method achieved a TDR of 100\%, with no false negatives or false positives, shown in Table~\ref{tab:CMFDreal}. In comparison, the SoP approach used in the industry has TDR of 91.8\%. Moreover, the FAR is 0\% for the FD filter, and 4.2\% for the SoP approach, highlighting the superior performance of the proposed FD filter. The improved performance of the proposed filter on experimental data is due to clearer distinctions between healthy and faulty signals. Simulated data includes slightly dried nozzles, harder to distinguish due to noise, whereas experimental faults are larger and easier to detect.


\subsection{Fault isolation results}
\label{subsection:resFI}
To evaluate the overall performance of the FI filter, the Harmonic Mean Average (HMA) is calculated. HMA is a standard metric used to assess the overall effectiveness of fault isolation schemes \cite{vanhybrid}. The performance of the proposed hybrid model-data fault isolation method is compared to that of a purely data-driven approach. The latter applies the FI filter directly to $y$, while the proposed method applies it to $r$, where $r$ is derived from the FD filter that incorporates model information. Additionally, the effect of training dataset size is examined by reducing it by 50\% and 90\%.

The results of the proposed FI methods using LR and KNN for the simulated data are presented in Table~\ref{tab:FIsim_comp}. These results demonstrate that for both methods and all training sizes, applying the FI filter to the residual signal $r$ consistently outperforms applying it to the system output signal $y$, highlighting the value of model information. Among fault classes, intermediately dried nozzles have the lowest classification accuracy (see Table~\ref{tab:CMFIsim} for the confusion matrix corresponding to the method using $r$, LR and a training size of 2025). This is likely because their signals share similarities with those of both slightly dried nozzles and partially blocked nozzles. 

\begin{table}[t]
\centering
\caption{FI results (in terms of HMA) with different training sizes for the simulated data.}
\label{tab:FIsim_comp}
\begin{tabular}{|c|c||c|c|}
\hline
\begin{tabular}[c]{@{}c@{}}Training\\ size\end{tabular} & \begin{tabular}[c]{@{}c@{}}Signal\\ used\end{tabular} & LR & KNN \\ \hline \hline
\multirow{2}{*}
{2025} & $r$ & 81.3\% & 80.0\% \\ \cline{2-4} 
 & $y$ & 73.1\% & 74.1\% \\ \hline
\multirow{2}{*}{1013} & $r$ & 79.4\% & 80.1\% \\ \cline{2-4} 
 & $y$ & 70.5\%  & 72.1\% \\ \hline
\multirow{2}{*}{203} & $r$ & 80.1\%  & 78.8\% \\ \cline{2-4} 
 & $y$ & 61.4\%  & 66.4\% \\ \hline
\end{tabular}
\end{table}

\begin{table}[b]
\centering
\caption{Confusion matrix of the proposed FI filter results for the simulated data (see Section~\ref{section:SOPD} for the abbreviations).}
\label{tab:CMFIsim}
\begin{tabular}{cl|cccccc|}
\cline{3-8}
\multicolumn{1}{c}{} &  & \multicolumn{6}{c|}{FI filter} \\ \cline{3-8} 
\multicolumn{1}{c}{} &  & \multicolumn{1}{c|}{EC} & \multicolumn{1}{c|}{BN} & \multicolumn{1}{c|}{PBN} & \multicolumn{1}{c|}{SDN} & \multicolumn{1}{c|}{IDN} & DDN \\ \hline
\multicolumn{1}{|c|}{\multirow{6}{*}{Ground truth}} & EC & \multicolumn{1}{c|}{225} & \multicolumn{1}{c|}{0} & \multicolumn{1}{c|}{0} & \multicolumn{1}{c|}{0} & \multicolumn{1}{c|}{0} & 0 \\ \cline{2-8} 
\multicolumn{1}{|c|}{} & BN & \multicolumn{1}{c|}{0} & \multicolumn{1}{c|}{225} & \multicolumn{1}{c|}{0} & \multicolumn{1}{c|}{0} & \multicolumn{1}{c|}{0} & 0 \\ \cline{2-8} 
\multicolumn{1}{|c|}{} & PBN & \multicolumn{1}{c|}{0} & \multicolumn{1}{c|}{0} & \multicolumn{1}{c|}{190} & \multicolumn{1}{c|}{35} & \multicolumn{1}{c|}{0} & 0 \\ \cline{2-8} 
\multicolumn{1}{|c|}{} & SDN & \multicolumn{1}{c|}{0} & \multicolumn{1}{c|}{0} & \multicolumn{1}{c|}{0} & \multicolumn{1}{c|}{225} & \multicolumn{1}{c|}{0} & 0 \\ \cline{2-8} 
\multicolumn{1}{|c|}{} & IDN & \multicolumn{1}{c|}{0} & \multicolumn{1}{c|}{0} & \multicolumn{1}{c|}{103} & \multicolumn{1}{c|}{113} & \multicolumn{1}{c|}{9} & 0 \\ \cline{2-8} 
\multicolumn{1}{|c|}{} & DDN & \multicolumn{1}{c|}{0} & \multicolumn{1}{c|}{0} & \multicolumn{1}{c|}{0} & \multicolumn{1}{c|}{0} & \multicolumn{1}{c|}{1} & 224 \\ \hline
\end{tabular}
\end{table}

For comparison, the manufacturer's SoP method achieves an HMA of 45.3\% on the largest training dataset. However, the SoP method has not been trained for all fault classes and would need expansion for a fair comparison.

The FI filter is also evaluated using experimental data, with results presented in Table~\ref{tab:FIreal_comp}.
\begin{table}[t]
\centering
\caption{FI results (in terms of HMA) with different training sizes for the experimental data.}
\label{tab:FIreal_comp}
\begin{tabular}{|c|c||c|c|}
\hline
\begin{tabular}[c]{@{}c@{}}Training\\ size\end{tabular} & \begin{tabular}[c]{@{}c@{}}Signal\\ used\end{tabular} & LR & KNN \\ \hline \hline
\multirow{2}{*}{147} & $r$ & 99.0\% & 99.0\% \\ \cline{2-4} 
 & $y$ & 98.0\% & 98.0\% \\ \hline
\multirow{2}{*}{74} & $r$ & 100\% & 100\% \\ \cline{2-4} 
 & $y$ & 98.0\% & 98.0\% \\ \hline
\multirow{2}{*}{15} & $r$ & 99.0\% & 99.0\% \\ \cline{2-4} 
 & $y$ & 98.0\% & 98.0\% \\ \hline
\end{tabular}
\end{table} 
For this dataset, all methods (whether using $r$ or $y$ for FI and regardless of training set size) achieved high accuracy due to the smaller number of fault classes compared to the simulated data. The SoP method's HMA is 88.3\%, but its primary limitation was 
misclassifying intermediately dried nozzles as healthy. This issue does not occur with the proposed FI filter, which significantly outperforms the SoP method.

It is worth emphasizing that the proposed method offers additional advantage. Unlike the SoP method, which performs FD and FI simultaneously, the proposed method isolates faults only after detecting them, achieved by leveraging model information in the design of the FD filter and using data for the FI filter, whereas the SoP method relies solely on data. 

\section{Conclusion} 
\label{section:conclusion}
This paper has presented a method for detecting and isolating faults in the ink channels of industrial printers using model-based detection filters and linear regression- and KNN-based fault classifiers. Using residuals (generated for fault detection) in fault isolation consistently outperformed directly using system output data, highlighting the advantages of the proposed method over purely data-driven approaches. Experimental results confirmed that the proposed fault isolation (FI) filter outperformed the state-of-the practice method, particularly by reducing the misclassification of faulty signals as healthy. For future work, it is recommended to experimentally evaluate the performance of the proposed methods with additional fault scenarios such as an air bubble in the ink channel.



\section*{Acknowledgement}
This publication is part of the project Digital Twin project 4.3 with project number P18-03 of the research programme Perspectief which is (mainly) financed by the Dutch Research Council (NWO). The last author also acknowledges the support of the European Research Council (ERC) under the grant TRUST-949796. The authors were enabled by Canon Production Printing Netherlands B.V. to perform research that partly forms the basis for this paper. Canon Production Printing B.V. does not accept responsibility for the accuracy of the data, opinions and conclusions mentioned in this paper, which are entirely for the account of the authors.

\appendix

\section{Identified system matrices}
\label{app:sysparams}
The system identification results are given as follows: 
\begin{equation*}
\begin{aligned}
    A &= \begin{bmatrix}
        0 & 0 & 1 & 0 \\
        0 & 0 & 0 & 1 \\
        -4.33 \times 10^{11} & -4.33 \times 10^{11} & -1.59 \times 10^{5} & 0 \\
        -6.17 \times 10^{11} & -6.17 \times 10^{11} & 0 & -1.75 \times 10^{5}
    \end{bmatrix}, \\
    C &= \begin{bmatrix}
        0 & 0 & 1 & 1
    \end{bmatrix},\quad 
    \hat{x}_a = 
    \begin{bmatrix}
        0.21 & 0.21 & 0.16 & 0.22
    \end{bmatrix}^\top.
\end{aligned}
\end{equation*}
\bibliographystyle{unsrt}
\bibliography{library/references}

\end{document}